\documentclass[aps,prl,twocolumn,superscriptaddress,showpacs,preprintnumbers]{revtex4-1}
\usepackage[colorlinks=true, pdfstartview=FitV, linkcolor=red, citecolor=blue, urlcolor=black, pdftitle={},pdfauthor={Tomoya Hayata},pdfsubject={}, pdfkeywords={}]{hyperref}
\usepackage{graphicx}
\usepackage{setspace,bm}
\usepackage{latexsym,amssymb,amsmath,mathrsfs}
\usepackage{color}
\graphicspath{{./figure/}}

\makeatletter
\providecommand \@ifxundefined [1]{%
 \@ifx{#1\undefined}
}%
\providecommand \@ifnum [1]{%
 \ifnum #1\expandafter \@firstoftwo
 \else \expandafter \@secondoftwo
 \fi
}%
\providecommand \@ifx [1]{%
 \ifx #1\expandafter \@firstoftwo
 \else \expandafter \@secondoftwo
 \fi
}%
\providecommand \bibnamefont  [1]{#1}%
\providecommand \bibfnamefont [1]{#1}%
\providecommand \citenamefont [1]{#1}%
\providecommand \href@noop [0]{\@secondoftwo}%
\providecommand \href [0]{\begingroup \@sanitize@url \@href}%
\providecommand \@href[1]{\@@startlink{#1}\@@href}%
\providecommand \@@href[1]{\endgroup#1\@@endlink}%
\providecommand \@sanitize@url [0]{\catcode `\\12\catcode `\$12\catcode
  `\&12\catcode `\#12\catcode `\^12\catcode `\_12\catcode `\%12\relax}%
\providecommand \@@startlink[1]{}%
\providecommand \@@endlink[0]{}%
\providecommand \url  [0]{\begingroup\@sanitize@url \@url }%
\providecommand \@url [1]{\endgroup\@href {#1}{\urlprefix }}%
\providecommand \urlprefix  [0]{URL }%
\providecommand \Eprint [0]{\href }%
\providecommand \doibase [0]{http://dx.doi.org/}%
\providecommand \selectlanguage [0]{\@gobble}%
\providecommand \bibinfo  [0]{\@secondoftwo}%
\providecommand \bibfield  [0]{\@secondoftwo}%
\providecommand \BibitemOpen [0]{}%
\providecommand \EOS [0]{\spacefactor3000\relax}%
\providecommand \BibitemShut  [1]{\csname bibitem#1\endcsname}%
\let\auto@bib@innerbib\@empty

\newcommand{\cE}{{\cal E}}
\newcommand{\cB}{{\cal B}}

\newcommand{\ve}{\varepsilon}

\newcommand{\be}{\begin{equation}}      
\newcommand{\ee}{\end{equation}}      
\newcommand{\bea}{\begin{eqnarray}}      
\newcommand{\eea}{\end{eqnarray}}

\begin{document}
\title{A new route to negative refractive index from topological metals}

\author{Tomoya Hayata}
\email[]{hayata@phys.chuo-u.ac.jp}
\affiliation{
Department of Physics, Chuo University, 1-13-27 Kasuga, Bunkyo, Tokyo, 112-8551, Japan 
}

\date{\today}

\begin{abstract}
We theoretically discuss the possibility of realizing the negative refractive index in Weyl/Dirac semimetals.
We consider the Maxwell equations with the plasma gap and the chiral magnetic effect.
We study the dispersion relations of electromagnetic waves, and show that the refractive index becomes negative at frequencies (just) below the plasma frequency.
We find that axial anomaly, or more specifically, negative magnetoresistance (electric current parallel to magnetic fields) opens a new route to realize the negative refractive index.
Reflection and transmission coefficients are computed in a slab of Weyl/Dirac semimetals.

\end{abstract}

\maketitle

{\it Introduction.} 
Negative refraction is the phenomenon that a light is counterintuitively refracted with a negative refractive angle compared to a natural matter. 
A matter exhibiting the negative refraction is refereed to as the negative-index material since its refractive index indeed becomes negative. 
Negative-index material has attracted a great deal of attention since it can realize the exotic phenomena never realized in a conventional material with the positive refractive index such as the perfect lens~\cite{PhysRevLett.85.3966,Fang534}.

Negative refraction was originally discussed by Veselago in Ref.~\cite{Veselago}. 
He required the permittivity and permeability to be negative simultaneously, and thus the material is sometimes called the double-negative material.
Such double-negative material has been experimentally realized by constructing an array of thin metal wires, and split ring resonators~\cite{Shelby77}.
After the first demonstration, various types of negative-index materials have been investigated, and the negative refractive index has been confirmed in broad regions from micro waves to optical wavelengths~\cite{Shalaev}.
Later, another mechanism not to require the double-negative permittivity and permeability has been indicated by using magnetoelectric effect, that is, cross polarization (magnetization) induced by magnetic fields (electric fields) in chiral materials~\cite{Tretyakov,Pendry1353,PhysRevLett.95.123904}, and confirmed in experiments~\cite{PhysRevLett.102.023901}.
The new mechanism of the negative refractive index is still interesting subject, since it has a chance to inspire broad device applications.

In this Letter, we theoretically discuss the possibility of realizing the negative refractive index in topological (semi-)metals such as Weyl/Dirac semimetals.
We study the Maxwell equations in the presence of the chiral magnetic effect (CME)~\cite{Fukushima:2008xe} with the chiral chemical potential dynamically generated by DC parallel electric and magnetic fields~\cite{PhysRevB.88.104412}.
By analyzing the dispersion relations of electromagnetic waves, we show that the interplay between CME and the plasma gap leads to the negative refractive index for either of right- or left-handed polarization at frequencies lower than the plasma frequency.
We show that axial anomaly, or more specifically, negative magnetoresistance (electric current parallel to magnetic fields) opens a new route to realize the negative refractive index.
Finally, transmission coefficients in a slab of Weyl/Dirac semimetals are computed, which can be used to retrieve the negative refractive index in actual experiments~\cite{PhysRevB.65.195104}.
We use natural units in this Letter.

{\it Chiral magnetic effect.} 
We start with a brief review on CME. 
In Weyl or Dirac fermions with nonzero chiral chemical potential $\mu_5$, static or dyamic magnetic fields $\bm B$ induces electric current because of (static) CME~\cite{Fukushima:2008xe}:
\bea
\bm j_{\rm sCME}=\sigma_{\rm sCME}\bm B ,
\label{eq:CME}
\eea
where the coefficient $\sigma_{\rm sCME}=e^2\mu_5/2\pi^2$ is protected by axial anomaly.
For concreteness, we considered a single Dirac fermion.
In condensed matter materials named Weyl/Dirac semimetals, Weyl or Dirac fermions exist as low energy-excitations, and the chiral chemical potential $\mu_5$, or equivalently, the chiral number density $\rho_5$ can be dynamically generated by applying uniform DC electric and magnetic fields ($\bm \cE$ and $\bm \cB$) via the interplay between the chirality relaxation and the axial anomaly:
\bea
\frac{d\rho_5}{dt}=-\frac{\rho_5}{\tau}+\frac{e^2}{4\pi^2}\bm \cE\cdot\bm \cB ,
\eea
where $\tau$ is a phenomenological relaxation time of chirality.
If we consider three-dimensional massless Dirac fermion or a pair of right- and left-handed Weyl fermions, the chiral chemical potential $\mu_5$ is given as~\cite{Fukushima:2008xe}
\bea
\mu_5=\frac{3}{T^2+3\mu^2/\pi^2}\rho_5 ,
\eea
where we assumed temperature $T$ or chemical potential $\mu\gg \rho_5^{1/3}$.
Then in a steady state ($\frac{d\rho_5}{dt}=0$), we have~\cite{Li:2014bha}
\bea
\mu_5=\frac{3}{T^2+3\mu^2/\pi^2}\frac{e^2\tau}{4\pi^2}\bm \cE\cdot\bm \cB .
\label{eq:mu5}
\eea
When $\bm B$ in Eq.~\eqref{eq:CME} is identical to DC magnetic fields $\bm\cB$, the dissipative current~\eqref{eq:CME} gives the negative and anisotropic $B^2$-term to the resistivity in weak magnetic fields satisfying $\omega_c\tau\ll1$ with $\omega_c$ being the cyclotron frequency~\cite{PhysRevB.88.104412}. 
This is the so called negative magnetoresistance, and used as a signal of Weyl/Dirac fermions in transport experiments~\cite{Li:2014bha,Huang:2015gy}.  
Below instead of the DC current, we consider the AC current induced by magnetic waves $\bm B$ with Eq.~\eqref{eq:CME} and $\mu_5$ generated by DC electric and magnetic fields. 
Namely, we discuss the propagation of electromagnetic waves in Weyl/Dirac semimetals under DC electric and magnetic fields. 

Furthermore, when magnetic fields $\bm B$ depend on time, additional contributions may appear from induction and magnetization currents: 
\bea
\bm j_{\rm tot}
&=&\bm j_{\rm sCME}+\dot{\bm P}+\nabla\times\bm M 
\notag \\
&=&\sigma_{\rm sCME}\bm B+2\sigma_{\rm GME}\bm B
\notag \\
&=&\sigma_{\rm CME}\bm B ,
\label{eq:dCME}
\eea
where $\sigma_{\rm GME}=-e^2(\mu_5-\ve_5)/6\pi^2$, and $\ve_5$ is the energy difference in a pair of Weyl nodes, which becomes nonzero in Weyl semimetals with inversion-symmetry breaking~\cite{2017arXiv170501111A}. 
Such additional currents are known as the gyrotropic magnetic effect (GME)~\cite{zhong2016gyrotropic} or the dynamic chiral magnetic effect. We here show $\sigma_{\rm GME}$ in the clean limit of isotropic system, but it has more complicated form in general cases such as multi-Weyl semimetals~\cite{Hayata:2017khl}.

\begin{figure*}[t]
 \includegraphics[width=1.\textwidth]{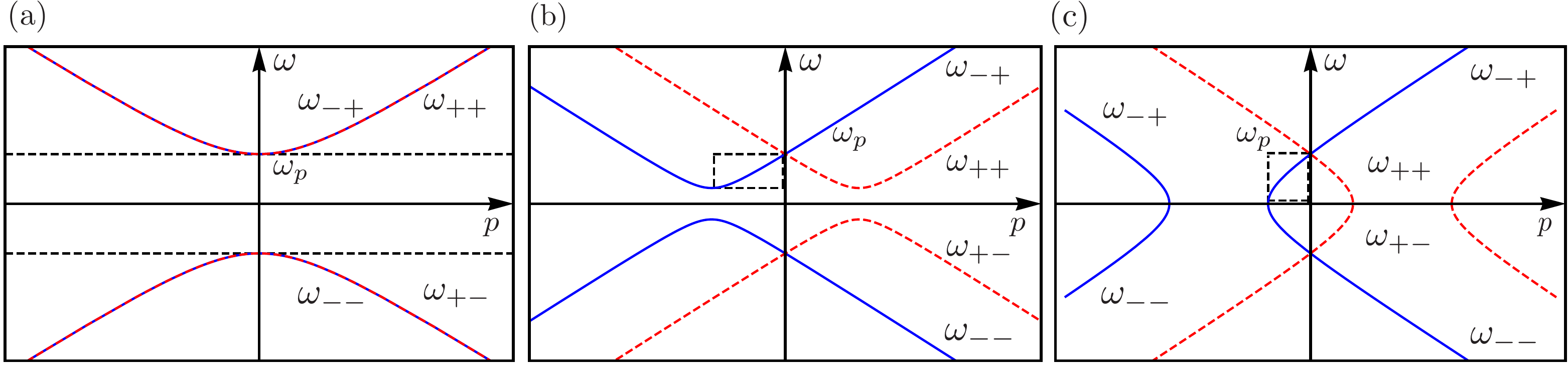}
 \caption{
Dispersion relations~\eqref{eq:dispersion} in three regimes: ($a$) $\sigma_{\rm CME}=0$, ($b$) $|\sigma_{\rm CME}|<2\omega_p$, and ($c$) $|\sigma_{\rm CME}|>2\omega_p$.
The first subscript on $\omega_{\chi\pm}$ refers to helicities, and the second the sign in Eq.~\eqref{eq:dispersion}.
When $\sigma_{\rm CME}\neq0$, the refractive index becomes negative in the regions indicated by dashed boxes.
The longitudinal mode is not shown in the figures, but it exists at $\omega=\omega_p$, and $\bm p=0$.
\label{fig1}
}
\end{figure*}

{\it Negative refractive index.} 
We discuss the propagation of monochromatic electromagnetic wave ($\bm E$ and $\bm B$) in Weyl/Dirac semimetals under uniform DC electric and magnetic fields ($\bm \cE$ and $\bm \cB$).
By taking CME~\eqref{eq:dCME} with $\mu_5$ in Eq.~\eqref{eq:mu5} into account, the Maxwell equations read
\bea
&&\nabla\cdot\left(\epsilon\bm E+\frac{i}{\omega}\sigma_{\rm GME}\bm B \right)=0 ,
\label{eq:Maxwell1}\\
&&\nabla \cdot\bm B=0 ,
\label{eq:Maxwell2}\\
&&\nabla\times \bm E =-\partial_t \bm B ,
\label{eq:Maxwell3}\\
&&\nabla \times \frac{1}{\mu_0}\bm B =\partial_t\epsilon \bm E+ \sigma_{\rm CME}\bm B  + \bm j_{\rm ext},
\label{eq:Maxwell4}
\eea
where $\bm E=\tilde{\bm E}e^{i\bm p\cdot\bm x-i\omega t}$, $\bm H=\tilde{\bm H}e^{i\bm p\cdot\bm x-i\omega t}$, and DC electric current $\bm j_{\rm ext}=\sigma_{\rm Ohm}\bm\cE+\sigma_{\rm sCME}\bm \cB$. 
We hereafter assume that the permittivity $\epsilon$, and permeability $\mu_0$ are isotropic.
We also assume the Drude-type permittivity: $\epsilon=\epsilon_0(1-\omega_p^2/\omega^2)$, with the plasma frequency $\omega_p$.
Below we take $\epsilon_0=\mu_0=1$ for simplifying expressions, but the generalization to arbitrary $\epsilon_0$ and $\mu_0$ is straightforward.
We neglected the chiral chemical potential generated by electromagnetic waves: $\delta\mu_5\sim-\tau(\bm E\cdot\bm\cB+\bm\cE\cdot\bm B)/(i\omega\tau-1)$ [$\bm E\cdot\bm B=0$]. 
This approximation is valid if $\omega\tau\gg1$ (the clean limit), or amplitudes of electromagnetic waves ($\bm E$ and $\bm B$) are small and nonlinear terms are negligible. 
The Maxwell equations are rewritten as
\bea
\left(\omega^2\epsilon-\bm p^2+\sigma_{\rm CME} \bm t\cdot\bm p \right)\bm E &=&0,
\label{eq:Maxwell5}
\\
\left(\omega^2\epsilon-\bm p^2+\sigma_{\rm CME} \bm t\cdot\bm p \right)\bm B &=&0 ,
\label{eq:Maxwell6}
\eea
where $i\bm p\times\bm E= (\bm t\cdot\bm p)\bm E$, and $[t_i]_{jk}=-i\epsilon_{ijk}$. 
The eigenvectors of Eqs.~\eqref{eq:Maxwell5}, and~\eqref{eq:Maxwell6} are helical polarizations satisfying
\bea
i\bm p\times\bm E_{\pm}= \bm t\cdot\bm p\;\bm E_{\pm}=  \pm p\bm E_{\pm} , \\
i\bm p\times\bm B_{\pm}=\bm t\cdot\bm p\;\bm B_{\pm}=  \pm p\bm B_{\pm} .
\eea
The dispersion relations are given explicitly as
\bea
\omega_{\chi\pm}=\pm \sqrt{\left(p_\chi- \chi\frac{\sigma_{\rm CME}}{2}\right)^2+\tilde{\omega}^2_p} ,
\label{eq:dispersion}
\eea
for right- ($\chi=+$) and left-handed helicities ($\chi=-$), respectively, and $\tilde{\omega}^2_p=\omega_p^2-\sigma_{\rm CME}^2/4$.
We show the dispersion relations~\eqref{eq:dispersion} in three regimes: ($a$) $\sigma_{\rm CME}=0$, ($b$) $|\sigma_{\rm CME}|<2\omega_p$, and ($c$) $|\sigma_{\rm CME}|>2\omega_p$ in Fig.~\ref{fig1}.
At $\sigma_{\rm CME}=0$, two polarizations are degenerate and show the gapped dispersion relations [See Fig.~\ref{fig1}($a$)]. The electromagnetic waves become evanescent at $\omega<\omega_p$ and cannot be transmitted except in thin films.
This is the usual characteristic of metals or plasmas.
On the other hand, in topological metals or plasmas, when $\sigma_{\rm CME}\neq0$, the helical degeneracy is lifted, and either of them has the negative refractive index [See Figs.~\ref{fig1}($b$) and ($c$)].

We can study the negative refractive index by two following ways.
We first focus on the sign of the group velocity $v_{g\chi}$ and phase velocity $v_{p\chi}$, which are given as
\bea
v_{g\chi}&=&\frac{\partial \omega_{\chi+}}{\partial p_\chi}
=\frac{p_\chi-\chi\frac{\sigma_{\rm CME}}{2}}{\sqrt{\left(p_\chi-\chi\frac{\sigma_{\rm CME}}{2}\right)^2+\tilde{\omega}^2_p}} ,
\label{eq:group}
\\
v_{p\chi}&=&\frac{\omega_{\chi+}}{p_\chi}
=\frac{\sqrt{\left(p_\chi-\chi\frac{\sigma_{\rm CME}}{2}\right)^2+\tilde{\omega}^2_p}}{p_\chi} .
\eea
Let us consider the regime ($b$) $|\sigma_{\rm CME}|<2\omega_p$.
In this case, the denominator in Eq.~\eqref{eq:group} is always positive, 
and $v_{g\chi}$ has the negative sign compared to $v_{p\chi}$ 
when $|p_\chi|<|\sigma_{\rm CME}|/2$, namely, when $\tilde{\omega}_p<\omega_\chi<\omega_p$ for either of helicities 
[See Fig.~\ref{fig1}($b$)].
This negative sign is the characteristic of the negative refractive index~\cite{Pendry1353}. 
Next we consider the regime ($c$) $|\sigma_{\rm CME}|>2\omega_p$. 
In this case, the dispersion relations~\eqref{eq:dispersion} become pure imaginary at some momentum range and indicate unstable modes~\cite{PhysRevLett.111.052002}. 
However, when we discuss transmission of electromagnetic waves,
the refractive index, that is, the (real) $\omega$-dependence of $p$ is used as information of dispersion relations, in which the unstable mode does not appear.
If we restrict ourselves to real $\omega$, gap just appears on momentum at 
$|p_\chi-\chi \sigma_{\rm CME}/2|<\sqrt{\sigma_{\rm CME}^2/4-\omega^2_p}$ 
[See Fig.~\ref{fig1}($c$)].
Below the gap, the denominator in Eq.~\eqref{eq:group} is positive, and $v_{g\chi}$ has the negative sign compared to $v_{p\chi}$ as seen in Fig.~\ref{fig1}($c$).
The characteristic of this regime is the lower bound on $\omega$ to have the negative refractive index vanishes, that is, all lights whose frequency is lower than $\omega_p$ have the negative refractive index.

We can explicitly see the negative refractive index by rewriting Eq.~\eqref{eq:dispersion} into the $\omega$-dependence of $p$.
Eq.~\eqref{eq:dispersion} is the quadratic equation on $p$, and we obtain the refractive index for each helicity from the positive sign solution as
\bea
n_{\chi}=\frac{p_{\chi+}}{\omega_\chi}
=\frac{\chi\frac{\sigma_{\rm CME}}{2}+\sqrt{\omega_\chi^2-\tilde{\omega}_p^2}}{\omega_\chi} .
\label{eq:refractive}
\eea
Let us again consider the regime ($b$) $|\sigma_{\rm CME}|<2\omega_p$.
When $\tilde{\omega}_p<\omega_\chi<\omega_p$,
we have $0<\sqrt{\omega_\chi^2-\tilde{\omega}_p^2}<|\sigma_{\rm CME}|/2$.
Therefore the refractive index~\eqref{eq:refractive} is negative for either of right- or left-handed helicity.
Next we consider the regime ($c$) $|\sigma_{\rm CME}|>2\omega_p$.
When $0<\omega_\chi<\omega_p$, 
we have $\sqrt{\sigma_{\rm CME}^2/4-\omega^2_p}<\sqrt{\omega_\chi^2-\tilde{\omega}_p^2}<|\sigma_{\rm CME}|/2$,
so that the refractive index~\eqref{eq:refractive} is negative for either of right- or left-handed helicity.
Since the numerator of Eq.~\eqref{eq:refractive} is nonzero at $\omega_\chi=0$, the refractive index~\eqref{eq:refractive} negatively diverges as $\omega_\chi\rightarrow0$.
Even in the regime ($b$), although the refractive index~\eqref{eq:refractive} is finite, it can take large negative values if $|\sigma_{\rm CME}|\sim2\omega_p$.
These behaviors might be important features of our mechanism to use CME.

Two remarks are in order: (I) The helicity of negatively refracted waves can be selected by changing the direction of DC electric and magnetic fields from parallel ($\bm \cE\cdot\bm \cB>0$) to anti-parallel ($\bm \cE\cdot\bm \cB<0$). 
This selectivity might be useful in a practical application.
(II) The frequency region to have the negative refractive index is determined by the plasma frequency $\omega_p$.
The magnitude of $\omega_p$ is estimated from the Fermi energy of Weyl or (gapless) Dirac nodes~\cite{PhysRevLett.102.206412}, 
and is typically $1$-$100$meV. 
Thus the negative refractive index originating from our mechanism might be relevant in THz region.

\begin{figure}[t]
\centering
 \includegraphics[width=.42\textwidth]{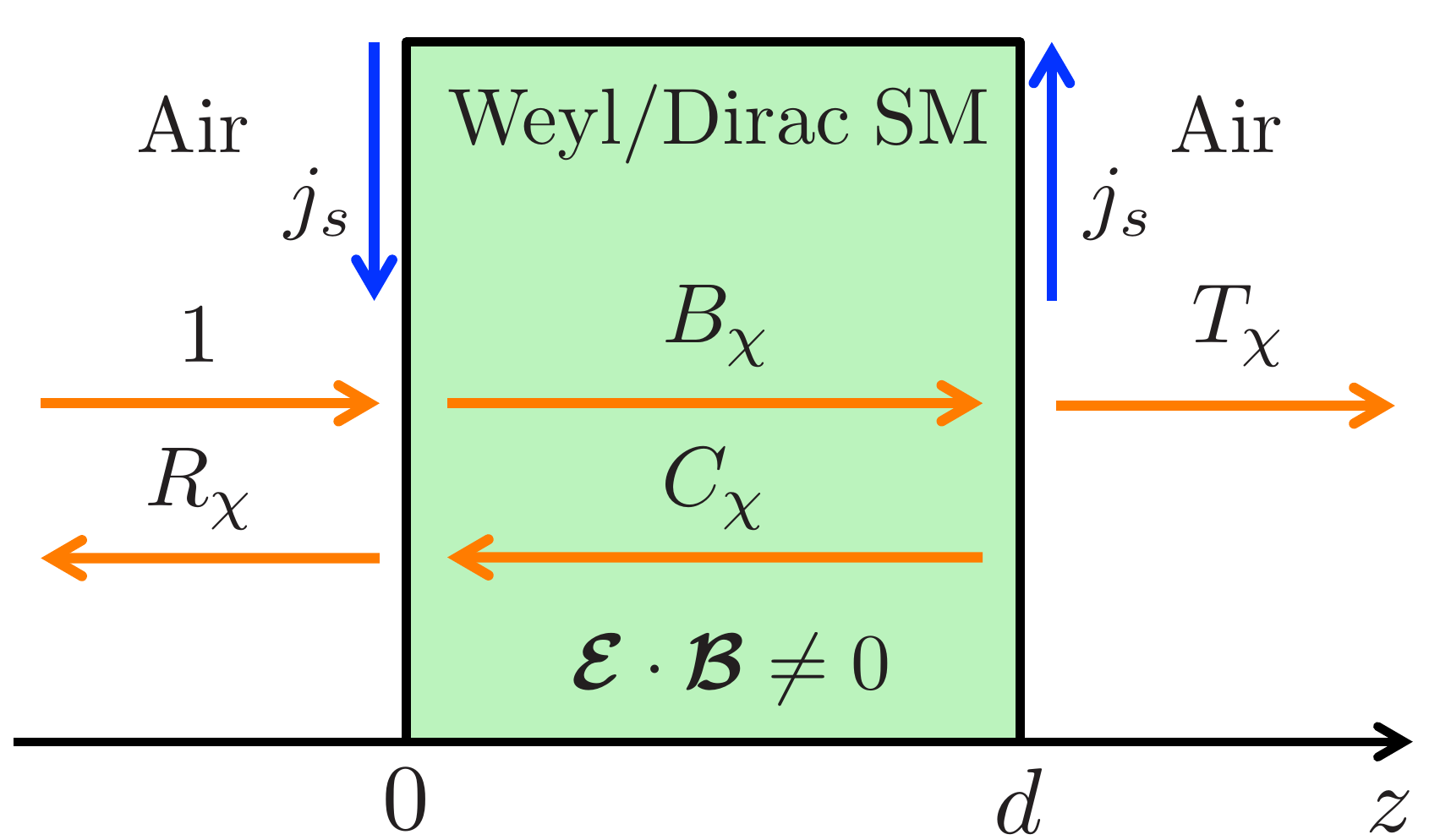}
 \caption{
Amplitudes of incident, reflected, and transmitted electric waves in a slab of Weyl/Dirac semimetal (SM).
The surface Hall current is induced by the incident electric waves.
\label{fig2}
}
\end{figure}

{\it Transmission coefficients.} 
We here study the transmission of electromagnetic waves in a finite slab of Weyl/Dirac semimetals with thickness $d$.
We consider right- and left-handed electromagnetic waves normally incident to the slab from the left as shown in Fig.~\ref{fig2}, and compute amplitudes of transmitted and reflected waves.
The transmission amplitudes can be used as experimental observables to retrieve the negative refractive index~\cite{PhysRevB.65.195104}.

By a similar computation with Ref.~\cite{PhysRevLett.115.117403} at $\sigma_{\rm CME}\neq0$~\cite{Hayata:2017khl}, 
we find the surface Hall current in the clean limit: 
\bea
\bm j_s=\mp \frac{i}{\omega}\left(\frac{\sigma_{\rm sCME}}{3}+\sigma_{\rm GME}\right)\hat{z}\times\bm E\delta(z-z_{\rm s}) ,
\label{eq:surfaec}
\eea
where the sign is $-$ for the front surface ($z_s=0$), and $+$ for the back surface ($z_s=d$), respectively.
We considered the isotropic Fermi-Dirac distribution at finite $T$, $\mu$, and $\mu_5$ as bulk distribution function to compute Eq.~\eqref{eq:surfaec}.
Considering the surface Hall current~\eqref{eq:surfaec}, the boundary condition on magnetic fields is written as 
$\hat{z}\times\left(\bm B_{\rm Air}-\bm B_{\rm SM}\right)=\hat{z}\times i(\sigma_{\rm sCME}/3+\sigma_{\rm GME})\bm E/\omega$.
Namely, the tangential components of $\tilde{\bm B}=\bm B+i(\sigma_{\rm sCME}/3+\sigma_{\rm GME})\bm E/\omega$, and $\bm E$ are continuous at boundaries.

The electric waves $\bm E_\chi=E_\chi\left(\hat{x}+i\chi\hat{y}\right)$ are given as
\bea
E_\chi=
\begin{cases}
e^{i\omega z}-R_\chi e^{-i\omega z} & \text{$z\leq0$} \\
B_\chi e^{in_\chi \omega z}-C_\chi e^{-in_{-\chi}\omega z} & \text{$0\leq z\leq d$}\\
T_\chi e^{i\omega(z-d)} & \text{$d\leq z$}
\end{cases} ,
\eea
where the amplitude of the incident wave is taken to be unity, and $\omega$ is the frequency at air.
The magnetic waves are obtained from the electric waves as $\bm B_\chi=\bm p\times\bm E_\chi/\omega$.
Then imposing the continuity conditions at $z=0$, and $z=d$ to $\bm E$, and $\tilde{\bm B}$ with the helical refractive index~\eqref{eq:refractive}, we obtain reflection and transmission coefficients: 
\bea
R_{\chi}&=&\frac{(n^2-\delta^2-1+2\chi\delta)\sin \kappa n}{2in\cos\kappa n+(1+n^2-\delta^2)\sin \kappa n} ,
\label{eq:reflection}
\\
T_\chi&=&\frac{2in e^{\chi i\sigma_{\rm CME}d/2}}{2in\cos\kappa n+(1+n^2-\delta^2)\sin \kappa n} ,
\label{eq:transmission}
\eea
where $n=(n_++n_-)/2=\sqrt{1-\tilde{\omega}_p^2/\omega^2}$, $\delta=\sigma_{\rm sCME}/6\omega$, and $\kappa=\omega d$.
When $\mu_5=0$ ($\delta=0$), but $\ve_5\neq0$, the time-reversal symmetry is recovered, and Eqs.~\eqref{eq:reflection}, and~\eqref{eq:transmission} reproduce those in gyrotopic medium~\cite{PhysRevB.73.045114}. 
From Eq.~\eqref{eq:transmission}, we obtain the azimuthal angle rotation, and ellipticity of transmitted waves as $\theta_{\rm T}=({\rm Arg}\;T_+-{\rm Arg}\;T_-)/2$, and $\eta_{\rm T}=(1/2)\sin^{-1}(|T_+|^2-|T_-|^2)/(|T_+|^2+|T_-|^2)$.
In our calculation, $\theta_{\rm T}=\sigma_{\rm CME} d/2$ becomes constant, and the direction of rotation is determined by the sign of $\bm \cE\cdot\bm \cB$ or $\ve_5$. 
The effect of CME can directly be seen from the azimuthal rotation.
The ellipticity is zero since $|T_+|=|T_-|$. 
$|T_+|$ or $|T_-|$ shows the resonant behaviors and perfect transmission if $\kappa n=\pi m$ with $m$ being some integers.
Such behaviors are the same as those in conventional metals.
Similarly, we can compute the azimuthal rotation, and ellipticity of reflected waves as $\theta_{\rm R}=0$, and 
\bea
\eta_{\rm R} &=&\frac{1}{2}\sin^{-1}\frac{4\delta(n^2-\delta^2-1)}{n^4+\delta^4-2n^2(1+\delta^2)+6\delta^2+1} 
\notag \\
&\sim& -\frac{\omega\sigma_{\rm sCME}}{3\omega_p^2},
\label{eq:ellipticity}
\eea
where we assumed $\omega\gg\sigma_{\rm sCME}$ ($\delta\ll1$) in the second line.
Eq.~\eqref{eq:ellipticity} indicates that we can directly access to the chiral chemical potential using electromagnetic waves.
Furthermore, combing $\theta_{\rm T}$ and $\eta_{\rm R}$, we can independently measure static and dynamic coefficients of CME ($\sigma_{\rm sCME}$, and $\sigma_{\rm GME}$).
Such an interesting optical property might come from the time-reversal symmetry breaking nature of the static CME or $\mu_5$, which can be nonzero only in nonequilibrium (steady) states~\cite{2017arXiv170501111A}.

{\it Summary.} 
In this Letter, we have studied the negative refractive index in Wey/Dirac semimetals.
We have analyzed the Maxwell equations in the presence of CME with the chiral chemical potential dynamically generated by DC electric and magnetic fields.
We show that the interplay between CME and the plasma gap leads to the negative refractive index for either of right- or left-handed polarization at frequencies lower than the plasma frequency.
Axial anomaly, or more specifically,  negative magnetoresistance (electric current parallel to magnetic fields)  can be used as a physical mechanism to realize the negative refractive index.
We have also computed reflection and transmission coefficients of right- and left-handed waves in a slab geometry for experimental implications.

The negative magnetoresistance (electric current parallel to magnetic fields) plays an essential role in our mechanism.
It has recently shown that the longitudinal negative magnetoresistance arises even in non-topological metals~\cite{PhysRevLett.120.026601}.
It will be interesting to study the negative refractive index in such non-topological metals.

Our analysis can directly be applied to relativistic plasma with nonzero chiral chemical potential.
It will be interesting to study phenomenological applications of the negative refraction to the physics of quark-gluon-plasma, and neutron stars such as magnetars.

\begin{acknowledgements}
The author thanks Y.~Hidaka for useful comments.
This work was supported by JSPS Grant-in-Aid for Scientific Research (No: JP16J02240).

Note added: After completion of this work, we found a similar work discussing the negative refraction in Weyl semimetals based on anomalous Hall effect~\cite{JPSJ.86.104703}.
Our proposal might be robust compared to Ref.~\cite{JPSJ.86.104703}, since it does not rely on microscopic structures of Weyl nodes, and can be applied to both of Weyl and Dirac semimetals. 
\end{acknowledgements}

\bibliography{./refractive}

\end{document}